\documentstyle[aps,prd,preprint,psfig]{revtex}
\begin{document}
\tightenlines

\def\stacksymbols #1#2#3#4{\def\theguybelow{#2}
    \def\verticalposition{\lower#3pt}
    \def\spacingwithinsymbol{\baselineskip0pt\lineskip#4pt}
    \mathrel{\mathpalette\intermediary#1}}
    \def\intermediary#1#2{\verticalposition\vbox{\spacingwithinsymbol
    \everycr={}\tabskip0pt
    \halign{$\mathsurround0pt#1\hfil##\hfil$\crcr#2\crcr
    \theguybelow\crcr}}}
    \def\lapproxeq{\stacksymbols{<}{\sim}{2.5}{.2}}
    \def\gapproxeq{\stacksymbols{>}{\sim}{3}{.5}}

\title{Open Universes, Inflation, and the Anthropic
Principle\footnote{Talk given at the XIV-th Yamada Conference, Kyoto,
April 1998.}}  \author{Alexander Vilenkin\footnote{Electronic address:
vilenkin@cosmos2.phy.tufts.edu}}

\address{Institute of Cosmology, Department of Physics and
        Astronomy,\\ Tufts University, Medford, Massachusetts 02155,
        USA} \date{\today} \maketitle

\begin{abstract}

Models of open inflation with a variable density parameter $\Omega$
provide the most natural way to reconcile an open universe with
inflation. The use of anthropic principle is essential to derive
observational predictions of such models. I discuss how this principle
can be used in a quantitative way to determine the most probable value
of $\Omega$. I also comment on recent work by Hawking and Turok on
quantum creation of open universes from nothing.

\end{abstract}

\section{Introduction.}

Until recently, thinking about open universes was regarded as a waste
of time because inflation predicted a flat universe. And thinking
about the anthropic principle could get you into real trouble. The
situation is now changing for open universes, as observations point to
low values of the matter density. To meet this observational
challenge, new models of inflation have been developed \cite{1,2,3,4}
in which the density parameter $\Omega$ can take a wide spectrum of
values.

I think the bad reputation of the anthropic principle is also largely
undeserved. If indeed we live in an open universe, then it is hard to
explain the observed value of $\Omega$ without using the anthropic
principle. The same applies to a universe with a nonzero cosmological
constant. Here I am going to show how anthropic principle can be used
in a quantitative way to determine the probability distribution for
$\Omega$. I will first give an overview of open inflation in Section
II. Then, in Section III, I will introduce my favorite version of the
anthropic principle, which I call the "principle of mediocrity". In
Section IV this principle will be applied to open inflation to obtain
the probability distribution for $\Omega$. In Section V, I will argue
that observers should not be surprised to find themselves living at an
epoch when the curvature is about to dominate. Finally, in Section VI,
I will comment on a somewhat related work by Hawking and Turok who
discussed quantum creation of open universes from nothing.

\section{Open Inflation.}

\subsection{One-field model.}

The simplest model of open inflation \cite{2,3} assumes a scalar field
$\phi$ with a potential $V(\phi)$ of a rather special form. This
potential is assumed to have a metastable false vacuum separated by a
barrier from a flat slow roll region leading to the true vacuum. The
first stage of inflation occurs when the field $\phi$ is stuck in the
false vacuum. Occasionally $\phi$ tunnels through the barrier with the
formation of a bubble. $\phi$ then slowly rolls towards the true
minimum of the potential, resulting in a second stage of inflation
inside the bubble.

The process of bubble nucleation is described \cite{CDL} by a compact,
$O(4)$ - invariant instanton which is a solution of Euclidean field
equations. The nucleation probability is ${\cal P} \sim e^{-S_E}$,
where $ S_E$ is the Euclidean action of the instanton. The evolution
of a bubble after nucleation is obtained by analytically continuing
the instanton to Lorentzian signature. It can be shown \cite{CDL} that
the bubble interior is isometric to an open Robertson-Walker
universe. The homogeneity and isotropy of the bubble spacetime is
ensured by the high symmetry of the instanton. Hence, in this case the
horizon problem is solved not by a large amount of inflation but by
the symmetry of the bubble nucleation process.

In the space between bubbles, the rate of expansion is very high, and
bubble collisions are very rare. We can therefore think of the bubbles
as isolated open universes. All of these universes have the same value
of $\Omega$ which is determined by the number of e-foldings of
slow-roll inflation, $N$. For observationally interesting values of
$\Omega$ we need $N \approx 60$.

This simple model demonstrates that inflation can indeed be reconciled
with a low-density universe, but it requires a substantial amount of
fine-tuning. The potential $V(\phi)$ is required to have a sharp
barrier next to a flat slow-roll region, which is a rather unnatural
combination.

\subsection{Two-field model.}

A more natural model, introduced by Linde and Mezhlumian \cite{4}, has
two fields, $\sigma$ and $\phi$, with $\sigma$ doing the tunneling and
$\phi$ the slow roll. The potential has the form
\begin{equation}
V(\sigma, \phi) = V_0(\sigma) + {1\over{2}}g\sigma^2\phi^2,
\label{1}
\end{equation}
where $V_0(\sigma)$ has a metastable false vacuum at $\sigma = 0$ and
the true vacuum at $\sigma = \sigma_0$. When $\sigma$ is in the false
vacuum, the field $\phi$ is massless, while in the true vacuum it has
a mass
\begin{equation}
m = g^{1/2}\sigma_0.
\label{2}
\end{equation}
A massless scalar field in an inflating universe is subject to quantum
fluctuations which can be pictured as a random walk along the
$\phi$-axis. After a while, all values of $\phi$ become equally
probable.

In this model bubbles can nucleate at different values of $\phi$. The
nucleation probability is
\begin{equation}
{\cal P} \sim e^{-S_E(\phi)}
\label{3}
\end{equation}
The number of e-foldings of the slow-roll inflation inside the bubbles
is also $\phi$-dependent,
\begin{equation}
N(\phi) \approx 2\pi G\phi^2.
\label{4}
\end{equation}
Hence, one expects to have a distribution of bubbles with different
values of $\Omega$ \cite{4}.

However, recent analysis by Garcia-Bellido, Garriga and Montes
\cite{5} has shown that this picture is oversimplified. The field
$\phi$ is not homogeneous inside the bubbles. The reason is that the
bubbles expand into the region of fluctuating field $\phi$, and the
fluctuations penetrate through the bubble wall. Mathematically, this
is described by the so-called supercurvature modes. Let us denote by
$t_\sigma$ the time it takes the field $\sigma$ to settle to its true
minimum at $\sigma_0$. (Here, $t$ is the Robertson-Walker time inside
a bubble). Then, on constant-time surfaces $t \sim t_\sigma$, $\phi$
has a gaussian distribution
\begin{equation}
{\cal P}(\phi) \propto \exp(-\phi^2/2<\phi^2>)
\label{5}
\end{equation}
with a dispersion \cite{5}
\begin{equation}
<\phi^2> \sim m^{-2}R_0^{-4}.
\label{6}
\end{equation}
Here, $R_0$ is the bubble radius at the time of nucleation. The
distribution (\ref{5}) is actually the same as that in Eq. (\ref{3}),
but their meanings are completely different. Eq. (\ref{3}) gives the
probability that a bubble nucleates with a given value of $\phi$,
while Eq. (\ref{5}) gives the probability distribution for $\phi$
inside a single bubble.

Only regions where $\phi$ is greater than the Planck mass $m_p$ are
going to inflate, and the amount of inflation will differ from one
region to another, according to Eq. (\ref{4}). Hence, each bubble will
contain an infinite number of regions with different values of
$\Omega$. Garcia-Bellido et. al. called this picture "quasiopen
inflation".

The correlation length of $\phi$ inside the bubbles at $t \sim
t_\sigma$ is $\xi \sim R_c/H_F^2m^2R_0^4$, where $H_F$ is the
expansion rate in the false vacuum and $R_c$ is the curvature radius
of the hypersurfaces $t \sim t_\sigma$. This correlation length sets
the length scale of variation of $\Omega$ and must be much greater
than the co-moving size of the presently observable universe (by a
factor of at least $10^7$, in order that the microwave background
anisotropies do not get unacceptably large). This can be enforced by a
suitable choice of parameters in the potential (\ref{1}).

Clearly, in this type of model, the value of $\Omega$ cannot be
predicted with certainty. One can only try to determine the
probability distribution for $\Omega$.

\section{Principle of Mediocrity.}

Each of the bubbles will be inhabited by an infinite number of
civilizations which will generally measure different values of
$\Omega$. In some regions $\Omega$ will be too low for any galaxies to
form. The probability for measuring such values of $\Omega$ should be
set equal to zero, since there will be nobody there to observe
them. When people talk about anthropic principle, they usually mean
this anthropic constraint (see, e.g., \cite{6}). However, I suggest
that we use a much more quantitative version \cite{7}.

My suggestion is that the probability ${\cal P}(\Omega)d\Omega$ for
$\Omega$ to be in the interval $d\Omega$ should be set proportional to
the number of civilizations which will measure $\Omega$ in that
interval. Assuming that we are a typical civilization, we can expect
to observe a value of $\Omega$ near the maximum of ${\cal
P}(\Omega)$. This version of the anthropic principle, which I called
the principle of mediocrity \cite{7}, is an extension of the
Copernican principle. The Copernican principle asserts that our
position in space is not special, while the principle of mediocrity
asserts that the values of the cosmological parameters we are going to
measure are not special either.\footnote{A very similar approach was
used by Carter \cite{8}, Leslie \cite{9} and Gott \cite{10} to
estimate the expected lifetime of our species. Gott also applied it to
estimate the lifetimes of various political and economic structures,
including the journal "Nature" where his article was published.}

A calculation of ${\cal P}(\Omega)$ based on the principle of mediocrity
was discussed in Ref. \cite{11}. At that time, the quasiopen nature of
the Linde-Mezhlumian model (\ref{1}) was not yet recognized. In the
next two sections I shall report on the recent paper I wrote with
Jaume Garriga and Takahiro Tanaka \cite{12} where we extend the work
of \cite{11} in two important respects. First, we use the quasiopen
picture (which actually makes the calculation much simpler) and
second, we give a much more careful treatment of the astrophysical
aspects of galaxy formation.

The principle of mediocrity has also been applied to other
cosmological parameters, e.g., the cosmological constant
\cite{7,efst,13,14,15} and the amplitude of density fluctuations \cite{16}.

\section{Calculation of ${\cal P}(\Omega)$.}

A great simplification introduced by the quasi-open picture is that
$\Omega$ takes all its possible values within each bubble. Since all
bubbles are statistically equivalent, it is sufficient to calculate
${\cal P}(\Omega)$ for a single bubble.

The distribution ${\cal P}(\Omega)$ can be expressed as
\begin{equation}
{\cal P}(\Omega) \propto {\cal
P}(\phi)e^{3N(\phi)}\nu(\Omega)\left|{d\phi\over{d\Omega}}\right|.
\label{7}
\end{equation}
Here, ${\cal P}(\phi)$ is the probability distribution for $\phi$,
Eq. (\ref{5}), and the next factor accounts for the fact that regions
starting out with different values of $\phi$ will inflate by a
different amount; $ e^{3N(\phi)}$ is the volume enhancement
factor. The "anthropic factor" $\nu(\Omega)$ is proportional to the
number of galaxies formed per unit volume. More precisely, it is the
fraction of galactic-scale volumes which eventually collapse to form
galaxies. (We assume that the number of civilizations is proportional
to the number of galaxies). Finally, the last factor in (\ref{7}) is
the Jacobian transforming from $\phi$ to $\Omega$ via the relation
\cite{2}
\begin{equation}
H_*^2e^{2N(\phi)} \approx {T_*^2\over{T_{eq}T_{CMB}}} \cdot
{\Omega\over{1 - \Omega}}.
\label{8}
\end{equation}
Here, $H_*$ and $T_*$ are the expansion rate and the temperature at
the end of inflation, $T_{eq}$ is the temperature at equal matter and
radiation densities, and $T_{CMB}$ is the temperature at the time when
the value of $\Omega$ is specified.

The Gaussian distribution ${\cal P}(\phi)$ is peaked at $\phi = 0$,
which corresponds to $\Omega = 0$, while the volume factor favors
large values of $\phi$ and pushes $\Omega$ towards $\Omega = 1$. The
effect of $\nu(\Omega)$ can be understood if we note that the growth
of density fluctuations in an open universe terminates at a redshift
$1 + z \sim \Omega^{-1}$. If $\Omega$ is too low, galaxy formation is
suppressed. Hence, $\nu(\Omega) \to 0$ as $\Omega \to 0$.

An interesting situation arises when ${\cal P}(\phi)$ dominates over
the volume factor. In this case, the effect of the anthropic factor
$\nu(\Omega)$ is to shift the peak of the distribution from $\Omega =
0$ to a non-zero value of $\Omega$.

In order to calculate $\nu(\Omega)$, one needs to make some
assumptions about the nature of cosmological density fluctuations. We
shall assume Gaussian fluctuations characterized by a dispersion
$\sigma_{rec}$ on the co-moving galactic scale at the time of
recombination. The choice of reference time here is unimportant, as
long as it is much earlier than the time of curvature domination, so
that one can reasonably assume that $\sigma_{rec}$ is independent of
$\Omega$. For the galactic scale we choose the co-moving scale of 1
Mpc at present.

In an open universe with a density parameter $\Omega_{rec}$ at
$t_{rec}$, the dispersion of density fluctuation in the asymptotic
future, $\sigma_\infty$, is greater than $\sigma_{rec}$ only by a
finite factor,
\begin{equation}
{\sigma_\infty\over{\sigma_{rec}}} = {5\over{2}}{\Omega_{rec}\over{1 -
\Omega_{rec}}}
 = {5\over{2}}{T_{rec}\over{T_{CMB}}}{\Omega\over{1 - \Omega}},
\label{9}
\end{equation}
where in the last step I have used the fact that
\begin{equation}
 T(1 - \Omega)/\Omega = const
\label{*}
\end{equation}
during the matter era. We can now use the Press-Schechter formalism to
determine $\nu(\Omega)$. Galaxies will form in regions where the
linearized density contrast $\delta$ exceeds the critical value
$\delta_c \approx 1.7$. Hence, $\nu(\Omega)$ is given by the integral
of the Gaussian distribution over $\delta > \delta_c$, that is, by the
error function \cite{17}
\begin{equation}
 \nu(\Omega) = {\rm
 erfc}\left({\delta_c\over{\sqrt{2}\sigma_\infty}}\right) = {\rm
 erfc}\left(\kappa{1 - \Omega\over{\Omega}}\right),
\label{10}
\end{equation}
where
\begin{equation}
\kappa =
{2\over{5}}{T_{CMB}\over{T_{rec}}}{\delta_c\over{\sqrt{2}\sigma_{rec}}}.
\label{11}
\end{equation}

In principle, the value of $\sigma_{rec}$ can be determined from the
fundamental theory of density fluctuations. Until such theory is
known, we can, in practice, adjust $\sigma_{rec}$ to fit the CMB
observations. Our ability to do so is, however, limited by the fact
that the value of $\sigma_{rec}$ inferred from observations depends on
the Hubble "constant" $H_0$ and the density parameter $\Omega_0$ in
our part of the universe, which are not very well determined. With
this uncertainty \cite{12},
\begin{equation}
\kappa = 0.1 \pm 0.05.
\label{12}
\end{equation}

Combining Eqs. (\ref{5}) - (\ref{8}) and (\ref{10}), we can now write
the final form of the probability distribution for $\Omega$,
\begin{equation}
{d{\cal P}\over{d \ln y}} \propto y^{3(\mu - {1\over{2}})}{\rm
erfc}(y).
\label{13}
\end{equation}
Here,
\begin{equation}
y = \kappa{1 - \Omega\over{\Omega}}
\label{14}
\end{equation}
and
\begin{equation}
\mu = {m_p^2\over{24\pi<\phi^2>}} \sim m_p^2m^2R_0^4.
\label{15}
\end{equation}
Note that the variable $y$ defined in (\ref{14}) does not depend on
the temperature $T_{CMB}$ at which $\Omega$ is measured, because of
Eq. (\ref{*}).

For $y > 1$, the error function can be approximated as
\begin{equation}
{\rm erfc}(y) \approx {1\over{\sqrt{2}y}}e^{-y^2}
\label{17}
\end{equation}
and the value of $\Omega$ at the peak of the distribution (\ref{13})
can be expressed analytically,
\begin{equation}
\kappa\left({1 - \Omega\over{\Omega}}\right)_{peak} \approx
\left({3\over{2}}\mu - {5\over{4}}\right)^{1/2}.
\label{18}
\end{equation}
The peak is rather broad, with a width
\begin{equation}
\Delta\left({1 - \Omega\over{\Omega}}\right) \sim 5.
\label{19}
\end{equation}
Interesting values of $\Omega_{peak}$, which are not too close to
either 0 or 1, are obtained in models with $\mu \sim 1$ (which can be
easily constructed). Further details of the calculations and the
results can be found in Ref. \cite{12}.

The moral of this analysis is that, given a particle physics model,
the probability distribution for $\Omega$ can be unambiguously
calculated from first principles.

\section{The cosmic age coincidence.}

The usual objection against models with $\Omega < 1$ is that it is
hard to explain why we happen to live at the epoch when the curvature
is about to dominate. That is, why
\begin{equation}
t_0 \sim t_c,
\label{20}
\end{equation}
where $t_0$ is the present time and $t_c$ is the time of curvature
domination. Observers at $t << t_c$ would find $\Omega \approx 1$,
while observers at $t >> t_c$ would find $\Omega << 1$. It appears
that one needs to be lucky to live at a time when $\Omega$ differs
substantially from both 0 and 1. Here I am going to argue that the
coincidence (\ref{20}) is not as surprising as it may first seem
\cite{12}.

>From the analysis in the preceding section, we can expect to have
\begin{equation}
t_c \sim t_G,
\label{21}
\end{equation}
where $t_G$ is the time of galaxy formation. Without the anthropic
factor $\nu(\Omega)$, the probability distribution for $\Omega$ is
peaked at $\Omega = 0$, which corresponds to $t_c \to 0$. The role of
$\nu(\Omega)$ is to push the peak to values of $\Omega$ such that the
curvature domination occurs soon after galaxies are formed, so that
$t_c \sim t_G$.

We now recall Dicke's observation \cite{18} that the present time
$t_0$ is likely to be comparable to the lifetime of a typical main
sequence star, $t_0 \sim t_\star \sim 10^{10}$ yrs. Moreover,
observationally the epoch of structure formation, when giant galaxies
were assembled, is at $z_G \sim 1 - 3$, or $t_G \sim 10^9 - 10^{10}$
yrs. Since, $t_G \sim t_\star \sim t_0$, it follows from (\ref{21})
that $t_c \sim t_0$.

The above argument is based on the coincidence
\begin{equation}
t_G \sim t_\star,
\label{22}
\end{equation}
which cannot be explained in the framework of our model. The time of
galaxy formation $t_G$ depends on the amplitude of the cosmological
density fluctuations, while the stellar lifetime $t_\star$ is
determined by the fundamental constants. In our model, the only
variable is $t_c$, while $t_G$ and $t_\star$ remain fixed. It is
conceivable that the coincidence (\ref{22}) may find some kind of
anthropic explanation in more general models where some other
"constants" are allowed to vary.

\section{Quantum creation of open universes.}

Hawking and Turok (HT) have recently argued \cite{19} that open
universes can be spontaneously created from nothing and suggested an
instanton to describe this process. The idea of using instantons to
describe the creation of the universe is not new \cite{20}. In the
case of a homogeneous closed universe, the corresponding instanton is
geometrically a four-dimensional sphere. In models with a metastable
false vacuum, the same Coleman-de Luccia instanton \cite{CDL} that is
used to describe bubble nucleation can be interpreted as describing
nucleation of a closed universe with a bubble \cite{21}. Analytic
continuation of this instanton gives a closed inflating universe with
an expanding bubble, the interior of the bubble being isometric to an
open Robertson-Walker space. In the course of the following evolution,
an infinite number of bubbles will nucleate in addition to the initial
one, and for all practical purposes it makes no difference whether the
universe nucleates with a bubble or without. So this sort of quantum
creation of open universes is not particularly interesting.

The new point of the HT paper is the claim that open universes can be
created even in models with very simple potentials, like
\begin{equation}
V(\phi) = {1\over{2}}m^2\phi^2,
\label{}
\end{equation}
which have no false vacua. Such models are known not to have regular
instanton solutions, and indeed, the HT instanton is
singular. Geometrically, it is like a sphere with a thorn, the tip of
the thorn being the singularity where the curvature and the scalar
field are infinite. HT point out, however, that the singularity is
integrable and the instanton action is finite. Analytic continuation
of this instanton gives a closed, singular spacetime (for a detailed
discussion of its structure see, e.g., \cite{22}). A part of this
spacetime, is isometric to an open Robertson-Walker universe. The
singularity has the form of an expanding singular bubble
\cite{23}. However, it never hits an observer in the Robertson-Walker
part of the universe, and HT argue that the singularity is therefore
not a problem \cite{24}.

HT instantons have a free parameter corresponding to the strength of
the singularity. As this parameter is varied, the density parameter
$\Omega$ of the open universe also changes, and HT use an anthropic
approach\footnote{The version of the anthropic principle employed by
HT is different from the principle of mediocrity I am using here. They
take the anthropic factor $\nu(\Omega)$ to be proportional to the
spatial density rather than to the number of observers who measure the
corresponding value of $\Omega$. I find this choice rather arbitrary,
and therefore hard to justify.} to find the most probable value of
$\Omega$.

I think there are serious problems with HT approach. We are interested
in instantons because, being stationary points of the Euclidean
action, they give a dominant contribution to the Euclidean path
integral. In a singular instanton, the field equations are not
satisfied at the singularity, and such an instanton is not, therefore,
a stationary point of the action. All singular instantons should
therefore be highly suspect, unless there is a good reason to believe
that the singularity is spurious.

Moreover, for the same model (\ref{}) I have constructed an
asymptotically-flat singular instanton \cite{23}. Geometrically, it
looks like a flat space with a thorn. The behavior of the fields near
the singularity is identical to that in the HT instanton and the
action is finite, so there is absolutely no reason to reject my
instanton if HT instanton is legitimate. The analytic continuation of
my instanton gives a flat space with a singular sphere which expands
at a speed close to the speed of light. If this were indeed a
legitimate instanton, then we would have to conclude that flat space
is unstable with respect to nucleation of singular bubbles. The
nucleation probability can be made very high by adjusting the strength
of the singularity, and since this strength is a free parameter, the
universe in this picture would have already been overrun by expanding
singular bubbles. Since this is in a glaring contradiction with
observations, we have to conclude that HT instanton, as it stands,
cannot be used to describe the creation of open universes.

An interesting recent development is the paper by Garriga \cite{25}
where he shows that an instanton of HT type can be obtained as a 4-d
reduction of a non-singular 5-d Kaluza-Klein instanton. He also found
non-singular 5-d analogue of my asymptotically flat instanton. The
strength of the singularity of the 4-d instantons obtained by
reduction is no longer arbitrary; it is fixed by the fundamental
constants of the theory. As a result, both HT-type and asymptotically
flat instantons have no free parameters. Garriga found that, for a
sufficiently large radius of the compactified dimension, the
probability of flat space decay is negligibly small. It appears that
in this case there are no objections against using Garriga's instanton
to describe the creation of open universes. However, as noted by
Garriga himself, since his instanton has no free parameters, the value
of $\Omega$ in the resulting universe is fixed. Anthropic
considerations discussed in Sections III-V do not apply to this model,
and certain amount of fine-tuning is required to obtain a non-trivial
value of $\Omega$ at the present time. It would be interesting if
models of this type could be constructed that would allow for a
continuous range of $\Omega$.

\section*{Acknowlegements}

I would like to thank the organizers of this meeting for their warm
hospitality. It was a pleasure to be back to Kyoto, particularly on
this happy occasion of celebrating Humitaka Sato's birthday and his
accomplishments.


\begin{thebibliography}{99}
\bibitem{1} J. R. Gott, Nature {\bf 295}, 304 (1982).
\bibitem{2} M. Bucher, A. S. Goldhaber and N. Turok, Phys. Rev. {\bf
D52}, 3314 (1995).
\bibitem{3} K. Yamamoto, M. Sasaki and T.Tanaka, Ap. J. {\bf 455}, 412
(1995).
\bibitem{4} A. D. Linde and A. Mezhlumian, Phys. Rev. {\bf D52}, 5538
(1995).
\bibitem{CDL} S. Coleman and F. De Luccia, Phys. Rev. {\bf D21}, 3305
(1980).
\bibitem{5} J. Garcia-Bellido, J. Garriga and X. Montes,
hep-ph/9711214.
\bibitem{6} S. Weinberg, Phys. Rev. Lett. {\bf 59}, 2607 (1987).
\bibitem{7} A. Vilenkin, Phys. Rev. Lett. {\bf 74}, 846 (1995).
\bibitem{8} B. Carter, unpublished.
\bibitem{9} J. Leslie, Mind {\bf 101.403}, 521 (1992).
\bibitem{10} J. R. Gott, Nature {\bf 363}, 315 (1993).
\bibitem{11} A. Vilenkin and S. Winitzki, Phys. Rev. {\bf D55}, 548
(1997).
\bibitem{12} J. Garriga, T. Tanaka and A. Vilenkin, astro-ph/9803268.
\bibitem{efst} G. Efstathiou, M.N.R.A.S. {\bf 274}, L73 (1995).
\bibitem{13} A. Vilenkin, in {\it Cosmological Constant and the
Evolution of the Universe}, ed. By K. Sato {\it et. al.}, Universal
Academy Press, Tokyo, 1996 (gr-qc/9512031).
\bibitem{14} S. Weinberg, in {\it Critical Dialogues in Cosmology},
ed. By N. Turok, World Scientific, Singapore, 1997.
\bibitem{15} H. Martel, P. R. Shapiro and S. Weinberg, Ap. J. {\bf
492}, 29 (1998).

\bibitem{16}

M. Tegmark and M. J. Rees, Ap. J. {\bf 499}, 526 (1998).

\bibitem{17} W. H. Press and P. Schechter, Ap. J. {\bf 187}, 425
(1974).
\bibitem{18} R. H. Dicke, Nature {\bf 192}, 440 (1962).
\bibitem{19} S. W. Hawking and N. G. Turok, hep-th/9802030.
\bibitem{20} A. Vilenkin, Phys. Lett. {\bf 117B}, 25 (1982).
\bibitem{21} R. Bousso and A. D. Linde, gr-qc/9803068.
\bibitem{22} W. Unruh, gr-qc/9803050.
\bibitem{23} A. Vilenkin, hep-th/9803084.
\bibitem{24} N. G. Turok and S. W. Hawking, hep-th/9803156.
\bibitem{25} J. Garriga, hep-th/9804106.
\end{thebibliography}
\end{document}